\documentclass[12pt]{article}

\begin{document}

\title{About the depletion profile  in nucleation kinetics  }

\author{Victor Kurasov}

\date{Victor.Kurasov@pobox.spbu.ru}

\maketitle

\begin{abstract}
The profile of vapor around the droplet growing in
the diffusion regime
with known moving boundary  have been
calculated and expressed in terms of
special functions.
\end{abstract}

The problem of determination of depletion zone
around the growing droplet is widely
investigated in modern publications,
meanwhile  the problem of nucleation
kinetics construction has been already
solved in \cite{kurasov}. Now this fact
in widely recognized  \cite{korsakov}
but still several attempts \cite{zuv}
were made to present some other (less
sophisticated) models of metastable
phase depletion. Since the approach of
stationary solution
 \cite{zuv} was strongly criticized in
\cite{lanl02} the new version
\cite{schekin} of the theory mainly by
the  same authors of \cite{zuv} takes
into account non-stationary effects.
Namely the mentioned publication
initiates this consideration.

One has to stress that there is no need
to find the rate of droplet growth, this
value has been determined with a great
accuracy. Results on growing interface
are widely published nowadays (see for
example Physica A) and can be found
elsewhere. So, it gave the opportunity
to present in \cite{kurasov} solution
based on given intensity of growth. This
solution is absolutely balanced and was
also discussed and refined in
\cite{lanl97}, \cite{lanl98},
\cite{lanl02}, \cite{vestnik_profile}.

Nevertheless the authors of
\cite{schekin} continued to construct
the solution of self consistent problem
of the growing droplet. They failed to
solve the precise problem and they
formulated some iteration procedure and
wrote the  solution  as an integral in
the first approximation. One has to note
that earlier a more general solution was
written in \cite{lanl02} in application
to nucleation, but in principle it is
well known \cite{eger}. In \cite{lanl02}
the general solution,  which
closes the question of obtaining the
profile on the base of the known moving
boundary, was given. Certainly, one can fulfill the
next step of this recurrent procedure
but it is absolutely useless to concrete
numerical results.

The profile of droplet was given  in
\cite{schekin} as a value proportional
to the integral
$$
I = \int_z^{\infty} \exp(-x^2) (1 -
\frac{z^2}{x^2}) dx
$$
Having corrected the evident misprint
(in the next paper \cite{last} the authors
of \cite{schekin} used the results of
\cite{schekin} with corrections, i.e. the right variant of formulas)
we can reduce it to
$$
I = \int_z^{\infty} \exp(-x^2) (1 -
\frac{z^2}{x^2})^{1/2} dx
$$
One can easily calculate this integral
which is equivalent to the integral
calculated in \cite{lanl98}.

Really
$$
\int_z^{\infty} \exp(-x^2) (1 - \frac{z^2}{x^2})^{1/2} dx
=
2 z^2  \Gamma(3/2)
\Psi(\frac{3}{2}, \frac{3}{2} ; z^2)
$$
Here $\Gamma$ is the Gamma function,
$\Psi $ is the Confluent Hypergeometric
function.

We have to note that in \cite{kurasov},
\cite{lanl98} the variable $z$ was
shifted on a radius of droplet, which
gives equivalent asymptotic expressions
far from the embryo, i.e. in the regions
important for nucleation. But

In \cite{schekin} none of publications
 \cite{lanl97},
\cite{lanl98}, \cite{lanl02},
\cite{vestnik_profile} have been
mentioned or cited in any way.

It is more than symbolic that in
\cite{schekin} the authors stopped at
uncalculated integral and present only
numerical pictures. Why was the integral
left uncalculated? To give the answer
one has to keep in mind the following
arguments:
\begin{itemize}

\item
Really, in \cite{kurasov}, which was also
unmentioned by authors of \cite{schekin}
the integral was also left uncalculated.

\item
Only in the little paper \cite{lanl98}
lying in lanl-archive the representation
via standard special functions was given.  This
representation is the most natural and
the absence of this representation in
\cite{kurasov} was the certain defect of
narration.

\item
The paper \cite{lanl98} was the only
paper which has no analogous variants in
Russian.

\item
One has also to take into account
 that all authors  of \cite{schekin} except H.Reiss
work at the same department of Physics
in St.Petersburg University, where the
author of  \cite{kurasov},
\cite{lanl97}, \cite{lanl98},
\cite{lanl02}, \cite{vestnik_profile}
works. Contrary to \cite{lanl98} the
papers \cite{lanl97}, \cite{kurasov}
have been published in the St.Petersburg
University Scientific Journal or have
been reflected there as Summaries.

\end{itemize}

As the result one can make
 further conclusions which
are so trivial that can be omitted here.


\begin{thebibliography}{99}


\bibitem{kurasov}
V. Kurasov,  Physica  A , vol. 226
(1996) p.117 - 136

V.B. Kurasov,  Deponirovano v VINITI
(Deponed in VINITI)
 5147-B  from
1.06.1989, 50p. (in russian)

V.B. Kurasov,  Deponirovano v VINITI
(Deponed in VINITI)
 2590-B95  from
19.09.1995, 47p. (in russian)


Kurasov V.,
 Explicit role of density profile in nucleation kinetics
  Second international workshop  NPT-2002.
  Book of abstracts, SPb, 2002, p. 39-40


\bibitem{schekin}
A.P.Grinin, A.K.Schekin, F.M.Kuni,
E.A.Grinina and H.Reiss, {\it Non Steady
profile of vapor  concentration around
the Growing Droplet: Balance of the
condensing Vapor and the Moving Boundary
of the Droplet}, In: Nucleation and
Atmospheric aerosols, 16-th
International Conference, Kyoto, Japan
 2004, Kyoto University Press, Editors
Mikio Kusahara and Markku Kulmala,
p.375-379

\bibitem{last}
A.P.Grinin, I.A. Zhuvikina, F.M.Kuni
 and H.Reiss, {\it Nearest-Neighbor Drop
 in Homogeneous Nucleation in a Supersaturated
 Vapor }, In: Nucleation and
Atmospheric aerosols, 16-th
International Conference, Kyoto, Japan
2004, Kyoto University Press, Editors
Mikio Kusahara and Markku Kulmala,
p.15-19

\bibitem{korsakov}
V. B.  Korsakov , R. A. Suris
Nucleation kinetics under low vapor
supersaturation, Second International
workshop "Nucleation and non-linear
problems in first order phase
transitions" (NPT'2002), Book of
abstracts, 1 July - 5 July 2002 , St.
Petersburg, Russia p. 68-69

\bibitem{zuv}
 Zuvikina I.A.,   Grinin A.P.,  Kuni F.M.,
 Stochastic regularities of new phase nuclei
 ebullition in first order phase transitions,  Second
International workshop "Nucleation and
non-linear problems in first order phase
transitions" (NPT'2002), Book of
abstracts, 1 July - 5 July 2002 , St.
Petersburg, Russia, p.54-57

\bibitem{eger}
V.F.Zaitsev,  A.D.Polianin, A handbook
on differential equations in partial
derivatives, Precise solutions, Moscow,
1996, 496 p.


Karsloy G., Eger D, Teploprovodnost'
tverduh tel (Heat conductivity of solid
bodies), Moscow, Nauka, 1964, 488 p.

\bibitem{lanl97}
V.Kurasov Preprint cond-mat@lanl.xxx.gov
get 9806342

Kurasov V.B., Vestnik
(Herald)St.Petersburg University, Seria
 4, 1999, Issue 2 (N 11) p. 80 -
83

\bibitem{lanl98}
V.Kurasov Preprint cond-mat@lanl.xxx.gov
get 9810022

\bibitem{lanl02}
V.Kurasov Preprint cond-mat@lanl.xxx.gov
get 0207772


\bibitem{vestnik_profile}


Kurasov V.B., Vestnik
(Herald)St.Petersburg University, Issue
3 (20), 2002, p. 19-31




\end{thebibliography}
\end{document}